\documentclass{ptephy_v1}

\preprintnumber{XXXX-XXXX} 

\usepackage{amsmath,amssymb,amsthm,graphicx}
\usepackage{color}
\usepackage{ulem}
\usepackage{arydshln}

\theoremstyle{definition}
\newtheorem{algorithm}{Algorithm}

\begin{document}

\title{A nonparametric method to assess significance of events in search for gravitational waves with false discovery rate}


\author{Hirotaka Yuzurihara}
\affil{
Institute for Cosmic Ray Research, The University of Tokyo,
Higashi-Mozumi 238, Kamioka-cho, Hida-shi, Gifu 506-1205 Japan
\email{yuzu@icrr.u-tokyo.ac.jp}
}

\author{Shuhei Mano}
\affil{The Institute of Statistical Mathematics, 10-3 Midori-cho, Tachikawa, Tokyo 190-8562, Japan}

\author{Hideyuki Tagoshi}
\affil{Institute for Cosmic Ray Research, The University of Tokyo, 5-1-5 Kashiwanoha, Kashiwa, Chiba 277-8582, Japan
}


\begin{abstract}
In this paper, we present a consistent procedure to assess the significance of gravitational wave events observed by laser interferometric gravitational wave detectors based on the background distribution of detection statistic. 
We propose a non-parametric method to estimate $p$-value. 
Based on the estimated $p$-values, 
we propose a new procedure to assess the significance of a particular event with 
$q$-value which is the minimum false discovery rate that can be attained when calling the event significant.
The $q$-value gives us a criterion on the significance of events 
which is different from $P_{\rm astro}$ which is used in the LIGO-Virgo analysis 
and in other analysis. 
The proposed procedure is applied to the 1-OGC and 2-OGC catalogs \cite{N18}\cite{N20}.
For most of the events which were claimed significant in \cite{N18} and \cite{N20}, 
we also obtain the same results. 
However, there are differences in the significance for several marginal events. 
Since the proposed procedure does not require any assumptions on signal and noise, 
it is very simple and straightforward. 
The procedure is also applicable to other searches for gravitational waves
whose background distribution of detection statistic is difficult to know.
\end{abstract}

\subjectindex{gravitational wave search, compact binary coalescence}

\maketitle

\section{Introduction}
\label{sec1}
The first gravitational wave event from binary black hole coalescence, GW150914, was observed by advanced LIGO detectors in the first observing run (O1) \cite{GW150914}.
After the first detection, tens of gravitational wave
events were reported \cite{LV18}.
During the second observing run (O2), 
the first gravitational waves from a binary neutron star coalescence,
GW170817 \cite{GW170817}, were observed by LIGO \cite{LIGO} and Virgo \cite{Virgo}.
The follow-up observations by electromagnetic telescopes identified the host
galaxy in NGC4993. The event strongly suggests the existence of radioactive decay
of rapid neutron-capture process \cite{L17}. The discovery of these events
has opened the gravitational wave astronomy.  
During the third observing run (O3), many candidates events were reported \cite{ref:GraceDB},
and four events have been published individually \cite{gw190425,gw190412,gw190814,gw190521}. 
Very recently, the GWTC-2 catalog which reports the gravitational wave signals from compact binary coalescences
during the first half of O3 observation were released \cite{GWTC-2}. 
In the coming years
the network of gravitational wave detectors 
consisting of two LIGO detectors, Virgo and KAGRA\cite{KAGRA} plans to perform
coincident observation runs. As the detectors' sensitivities improve
and observation time becomes longer,
we expect to observe more and more gravitational wave events.

In compact binary coalescence searches, we search for gravitational wave signals by maximizing the detection statistic over the template bank in a short time window. When the value of the detection statistic exceeds a given threshold, we record it as a trigger.
Accordingly, for a given threshold, as the observing time and the template bank becomes larger, the probability that false triggers produced by noise (false alarm probability) becomes larger. This is called the multiple comparisons problem.  
Several methods have been proposed to control the false alarm probability. 
The Bonferroni correction is one of the method (see Chapter 9 of \cite{LR08}).
However, these methods generally reduce the detection probability while controlling the false alarm probability.

Recently, the false discovery rate (FDR) was proposed to treat these problems 
(see Section~\ref{sec3} for the formal definition of the FDR). By
the author's knowledge, the first introduction of FDR to the gravitational
wave community was done by Baggio and Prodi \cite{BP05}, but the paper did not
discuss any actual problems. 
Recently, $P_{\rm astro}$ was introduced as a measure of true discovery of a particular event
\cite{FGMC15}.
In the recent catalog of gravitational waves from compact binary mergers \cite{LV18},
a candidate event is considered to have gravitational wave origin, if the false
alarm rate is less than one per 30 days and
the $P_{\rm astro}$, is larger than 0.5.

In this paper, we propose the use of {\it q-value} which is a measure of FDR. 
We present a consistent procedure to assess the significance 
of candidate events by using $q$-value. 
We first introduce a definition of the $p$-value by using the background distribution of the detection statistic. 
Then, we propose a new procedure to evaluate $q$-value of each event by extending the procedure proposed by Storey and Tibshirani \cite{ST03}. 
The original procedure by Storey and Tibshirani \cite{ST03} is not applicable for a search for gravitational waves from compact binary coalescences,
because it requires a complete list of $p$-value. 
However, in gravitational wave searches, a complete list of $p$-value is usually not available because we store only triggers whose detection statistic is larger than a certain threshold. 
We apply these procedure to the publically available results of the analysis, 1-OGC catalog and 2-OGC catalog by Nitz et al. \cite{N18, N20}, and evaluate the $q$-value of each candidate event. 
We compare the significance of each candidate event evaluated by using $P_{\rm astro}$. 
We find that we obtain almost consistent results on the significance of each candidate event. 
However, we also find that, although the conclusion on the significance may change 
depending on the threshold for $q$-value and  $P_{\rm astro}$, 
the conclusion on the significance of events can be different for marginally significant events. 
We find one such event in 2-OGC catalog. 

The main advantage of our procedure is that our procedure is 
completely nonparametric, namely, we do not assume any parametric model 
behind data. Our procedure can be applied to other gravitational wave searches.  
The evaluation of $p$-value in non-parametric way, the procedure to  evaluate $q$-value, 
estimation of $q$-value for the LIGO-Virgo O1 and O2 candidate events by using this procedure, 
all of these are new things in this paper.

This paper is organized as follows. 
In section \ref{sec2}, 
we discuss statistical hypothesis testing
in the search for gravitational waves from compact binary coalescences.
In section \ref{sec3}, we present a procedure to assess a significance of a particular event with a false discovery rate. 
In section \ref{sec4}, the proposed procedure is 
applied to the results of the analysis of the O1 data.
Section \ref{sec5} is devoted to a summary and discussion.

\section{Estimation of $p$-value}
\label{sec2}

We first introduce the statistical terminologies used in this paper.
The definitions of statistical terminology can be found in a standard textbook, such as \cite{LR08}.
By analyzing the data from gravitational wave detectors,
we obtain \textit{events} which have larger signal-to-noise ratio than a threshold.
Each event is classified as either \textit{signal} or \textit{noise}.
If the event is originated from a gravitational wave, it is called a signal. Otherwise, it is called a noise.
In the statistical literature, the noise model is called \textit{null hypothesis}
(in this paper, also called \textit{background})
and the signal model is called 
\textit{alternative hypothesis}.

In the analysis of gravitational waves from compact binary coalescences, 
event search is done by maximizing the detection statistic over the templates. 
The detection statistic is also maximized over time within a certain time length.

In statistical hypothesis testing, the $p$\textit{-value} of an event is
a measure of the significance of the event. It is the probability that the event
or rarer events occur under the null hypothesis. If the $p$-value of the event
is significantly small, the null hypothesis is rejected.
Let us consider statistical hypothesis testing of each event based on 
background distribution of detection statistic.

\subsection{A conventional $p$-value}

In the LIGO-Virgo O1 analysis, the following $p$-value was used \cite{A16, U16}
(see Appendix A for discussion on the derivation)
\begin{equation}
  p_{\rm conv}(\rho)=1-e^{-\mu(\rho)}, \qquad
  \mu(\rho)=\frac{n_{\rm bg}(\rho)}{t_{\rm bg}}t_{\rm obs},
  \label{p-value_conv}
\end{equation}
where $\rho$ is the detection statistic of a event.
In this paper, we call this the conventional $p$-value.
Here, $t_{\rm obs}$ and
$t_{\rm bg}$ are the time length of the analyzed data and the time length for the estimation of the background distribution,
respectively.
The estimation of the background data is usually generated by time-shifting 
data of different detectors \cite{U16}.
Moreover, $n_{\rm bg}(\rho)$ is the number of
noise events in the background data whose detection statistics are
equal to or larger than $\rho$. It is
\begin{equation}
  n_{\rm bg}(\rho)=\sum_{i=1}^{n_{\rm bg}(0)}1_{\{r_i\ge \rho\}},
  \label{nbg}
\end{equation}
where 
$r_i$ is the detection statistic of the $i$-th event in the background data,
$1_{\{\cdot\}}=1$ if $\{\cdot\}$ is true and 0 otherwise.
From the definition,
$n_{\rm bg}(0)$ is the total number of noise events in the background data.
Therefore,
$\mu(\rho)$ in Eq.(\ref{p-value_conv}) is
the mean of number of events whose detection statistics are more than or equal to
$\rho$. The ratio $n_{\rm bg}(\rho)/t_{\rm bg}$ is usually called
the \textit{false alarm rate} of the event whose detection statistic is $\rho$.

\subsection{Nonparametric estimation of $p$-value}

Now, we introduce a non-parametric method to estimate $p$-value. 
Let us assume the background distribution is continuous.
If we know the probability density function of
detection statistic under the null hypothesis, $f(r)$,
the $p$-value of an event whose detection statistic is $\rho$ is given by
\begin{equation}
  {p(\rho)}=\int_\rho^\infty f(r)dr=1-F(\rho), \qquad F(\rho)=\int_0^\rho f(r)dr.
  \label{p-value}
\end{equation}
In reality the background distribution is unknown, nevertheless, it can be estimated
non-parametrically (free from assumption of a parameterized distribution) by
using simulated background data. An estimator of the null distribution $F$ is
given by
\[
  \hat{F}(\rho):=\frac{1}{n_{\rm bg}(0)}\sum_{i=1}^{n_{\rm bg}(0)}
  1_{\{r_i\le \rho\}}.
\]
It is important to distinguish $F$ and $\hat{F}$. The former is the (unknown) true background distribution, while the latter is an estimator of the background distribution. By Glivenko-Cantelli's theorem, $\hat{F}$ converges to $F$ almost surely and uniformly in $\rho$ \cite{LR08}.
Therefore, an estimator of the $p$-value of an event whose the detection statistics is $\rho$ is given by
\begin{equation}
   {\hat{p}(\rho)}
   = 1-\hat{F}(\rho):=\frac{n_{\rm bg}(\rho)}{n_{\rm bg}(0)},
  \label{p-value_est}
\end{equation}
where we used the fact that $\rho\neq r_i$ where $i = {1,...,n_{\rm bg}(0)}$.
Note that $\hat{p}(\rho)$
is the probability of obtaining the event whose the detection statistics is larger than $\rho$ in the background data and
has been called (an estimator of)
\textit{false alarm probability}
in the gravitational wave community \cite{C17}.
In addition, $\hat{p}(\rho)$ is proportional to the mean $\mu(\rho)$ in (\ref{p-value_conv}).
The estimator (\ref{p-value_est}) is a consistent estimator of the $p$-value 
(\ref{p-value}), namely, $\hat{p}(\rho)$ converges to $p(\rho)$ almost surely
for each $\rho$ by the strong law of large numbers.

For later discussion, let us recall a basic property of a $p$-value.
A $p$-value of a statistic $\rho$ following any continuous
null distribution $F(\rho)$ follows the uniform distribution, because
\begin{align*}
  \mathbb{P}(p(\rho)<u)
  &=\mathbb{P}(\rho>F^{-1}(1-u))
  =1-\mathbb{P}(\rho \leq F^{-1}(1-u))\\
  &=1-F(F^{-1}(1-u))=u,
\end{align*}
is the distribution function of the uniform distribution where $0 \leq u \leq 1$
and  $\mathbb{P}(x)$ is the probability of $x$.
It is worthwhile to mention that we cannot expect that the {conventional} $p$-value given by (\ref{p-value_conv})
with $\rho$ following $F(\rho)$, follows the uniform distribution
{
(see Appendix A). In the discussion that follows, we discuss the $p$-value
defined by \eqref{p-value}.}

\section{Assessment of significance with false discovery rate}
\label{sec3}

In this section, we describe a statistical hypothesis testing by using detection statistics
and
how to assess a significance with the false discovery rate.
When we perform the statistical test,
each event can be categorized in
four possible outcomes, which are summarized in Table \ref{table1}.

\begin{table}[!hbt]
\begin{center}
\caption{Outcomes and counts}
\begin{tabular}{lccc}
					& Called significant & Called not significant & Total \\
\hline
Noise 	& $F$ & $n_0-F$ & $n_0$ \\
Signal 	& $T$ & $n_1-T$ & $n_1$ \\
Total  & $S$ & $n_{\rm obs}-S$   & $n_{\rm obs}$   \\
\hline
\end{tabular}
\label{table1}
\end{center}
\end{table}

There are two kinds of {\color{black}truth} (noise or signal) and two kinds of claim (called significant or called not significant).
$F$ and $T$ are the number of noise and signal events
called significant, respectively,
and $S$ is the total number of 
events called significant.
$n_0$ and $n_1$ are the number of noise and signal statistics, respectively. 
$n_{\rm obs}$ is the total number of events in the observed data.

In statistical hypothesis testing, a $p$-value threshold is
selected to keep the number of false positives $F$ small.
When we select the threshold $\alpha$, the expected number of false positive is $\alpha n_{\rm obs}$.
If $n_{\rm obs}$ is very large, $\alpha$
should be selected to be very small.

Here, the probability $\mathbb{P}(F\ge 1)$ is called a \textit{familywise error rate}.
The familywise error rate is simply called false alarm probability in the gravitational wave community,
but we call the familywise false alarm {probability} in
this paper to avoid a confusion.
The {\textit{family}} means that we test a hypothesis by using $n_{\rm obs}$ tests.
To control the familywise error rate such that $\mathbb{P}(F\ge 1)\le\alpha$, that is,
the rate that a noise event is classified as called significant is less than $\alpha$,
one of the solutions is to
change the threshold $\alpha$ to $\alpha/n_{\rm obs}$. 
This method is called Bonferroni's {procedure
  (see Chapter 9 of \cite{LR08}).}

Unfortunately, controlling the familywise error rate is practical only
when extremely few events are expected to be signal.
Otherwise, controlling the familywise error rate
will be too conservative and statistical power
of the test procedure will be too poor.
Benjamini and Hochberg \cite{BH95} introduce the \textit{false discovery rate},
which is defined as
the expected value of $F/S$, 
$\mathbb{E}(F/S,S>0)$, where $F$ and $S$ are introduced in Table \ref{table1},
and give a test procedure to keep the FDR less than a threshold.
A fairly recent survey of an FDR is \cite{B10}.
Note that the false positive rate and the FDR are quite different measures. A false positive rate
of $5\%$ means that $5\%$ of noise events are called significant. On the other
hand, an FDR of $5\%$ means that $5\%$ of events called significant are noise
events.
Controlling FDR should
be more powerful
than controlling 
familywise error rate,
since FDR is less than or equals to the
familywise error rate \cite{BH95}.

Storey and Tibshirani \cite{ST03} introduced
the \textit{$q$-value} for a particular event, which is the expected
proportion of false positives incurred if calling the event significant. Let
us define FDR$(u)$, which is the FDR when calling all events significant whose
$p$-value is less than or equals to a threshold $u$ where $0 < u \leq 1$,
namely,
\begin{equation}
  \text{FDR}( {\color{black}u} )=\mathbb{E}\left[\frac{F( {\color{black}u})}{S( {\color{black}u} )}, S( {\color{black}u} )>0
    \right],
\label{FDRt}
\end{equation}
where
$\mathbb{E}(x,y>0)$ is the expectation of $x$ given $y>0$.
Here, $F(u)$ is the number of the noise events whose $p$-value is smaller than or equals to the threshold ${\color{black}u}$, and 
$S( {\color{black}u} )$ is the number of both noise and signal events whose $p$-value is smaller than or equals to the threshold ${\color{black}u}$.
The definition of the $q$-value is the minimum FDR that can be attained when
calling the event significant, namely,
\begin{equation}
  q_i:=\min_{ {\color{black}u} \ge p_i} \text{FDR}( {\color{black}u} ),
  \label{q-value}
\end{equation}
where $i = 1,...,n_{\rm obs}$ and
the $p$-value given by (\ref{p-value}) of the $i$-th event are denoted by $p_i$.
Note that FDR$( {\color{black}u} )$ is not always monotonically
increasing in the threshold ${\color{black}u}$. Taking
minimum guarantees that the estimated $q$-value is increasing in the same
order as the $p$-value.

Let us recall the procedure for estimating  $q$-value proposed by Storey and
Tibshirani \cite{ST03}. Their estimator of the FDR$( {\color{black}u} )$ is
\begin{equation}
  \widehat{\text{FDR}}( {\color{black}u} )=\frac{\hat{\pi}_0 n_{\rm obs} {\color{black} u}}{S( {\color{black}u} )},
  \label{FDRt_est}
\end{equation}
where $\hat{\pi}_0$ is an estimator of $\pi_0=n_0/n_{\rm obs}$
which indicates the overall proportion of noise events in the data.
Roughly speaking, (\ref{FDRt_est}) is a sample mean whose population mean is (\ref{FDRt}).
Since a $p$-value of a statistic follows the uniform distribution under
the null hypothesis (see Section \ref{sec2}), the numerator of (\ref{FDRt_est}) is an estimator of $F( {\color{black}u} )$. 

How to estimate $\hat{\pi}_0$ is the central issue. 
In the gravitational wave searches, very few events are expected to be signal. 
In such a case, we can assume $\hat{\pi}_0 \simeq 1$. 
In Appendix \ref{appendixpi0}, we show that this assumption 
is justified by using the 1-OGC and 2-OGC catalogs. 
We thus set  $\hat{\pi}_0 = 1$.

We can construct
an estimator of the $q$-value by plugging the estimator of the $p$-value
(\ref{p-value_est}) and the estimator of the FDR (\ref{FDRt_est}) into
the expression (\ref{q-value}) and setting $\hat{\pi}_0=1$. The result is 
\begin{equation}
    \hat{q}_{i}
    =\min_{{\color{black} u} \ge \hat{p}_{i}}\frac{n_{\rm obs} {\color{black} u}}
    {\#\{\hat{p}_{j} \le u;j\in\{1,...,n_{\rm obs}\}\}},
    \label{q-value_est}
\end{equation}
where $\hat{p}_i=\hat{p}(r_i)$.

\section{Application to 1-OGC and 2-OGC results}
\label{sec4}
In this section, we evaluate $q$-value of events in the 1-OGC catalog  \cite{N18} 
and in the 2-OGC catalog  \cite{N20}.
We use the data available at {\tt https://github.com/gwastro/1-ogc} and {\tt https://github.com/gwastro/2-ogc}. 
Available data set contains the information of events such as time, false alarm rate in a unit of year$^{-1}$, 
the value of ranking statistic, two masses, dimensionless spin component value of each star perpendicular to the orbital plane, etc. 
The data set consists of \textit{complete} and \textit{bbh} data sets. 
There are 146,214 and 12,741 events in \textit{complete} and \textit{bbh} data sets of 1-OGC, and 
733,231 and 502,994 events in \textit{complete} and \textit{bbh} data sets of 2-OGC, respectively.  
The complete data set contains all candidate events from full analysis, and the bbh data set contains 
the candidate events from the BBH region targeted analysis \cite{N18, N20}. 

Since $p$-value of events are not available in these catalog, 
we need to evaluate it from the false alarm rates (FAR). 
An estimate of FAR is given by $n_{\rm bg}(\rho)/t_{\rm bg}$ where $t_{\rm bg}$ is the length of data used for background estimation, and $n_{\rm bg}(\rho)$ is defined by Eq. (\ref{nbg}).  
The events in the catalog are defined by taking an event which gives a maximum detection statistic within a certain time window $\Delta t$
and in the template bank used in the analysis. 
Thus, the total number of background, $n_{\rm bg}(0)$, is given as $t_{\rm bg}/\Delta t$. 
In both 1-OGC and 2-OGC, $\Delta t=10$ seconds are used. 
Then, from Eq. (\ref{p-value_est}), we obtain an estimate of $p$-value of 
an event as 
\begin{equation}
\hat{p}(\rho)=\frac{n_{\rm bg}(\rho)}{n_{\rm bg}(0)}
={\rm FAR} \times \Delta t.
\end{equation}

We note that the candidate events in these data sets are not all events in the sense that 
only events with relatively low false alarm rates are recorded. 
This is due to a practical reason in order to reduce the computation time of the analysis. 
This is a typical situation in gravitational wave analysis. 
 
Since all candidate events are not available, 
we can not use the algorithm originally proposed in \cite{ST03}, 
which is explained as Algorithm \ref{alg1} in Appendix \ref{sec:algorithm}. 
Instead, we propose an alternative procedure for estimating $q$-value 
which is a modified version of Algorithm~\ref{alg1}. 
Appendix \ref{sec:algorithm} explains why Algorithm~\ref{alg2} yields estimates
of the $q$-value defined in (\ref{q-value_est}).

\begin{algorithm}\label{alg2}

We compute estimates of $q$-value defined in (\ref{q-value_est}).
Let $m$ to be the number of false alarm rates which are less than some
value. Assume $p$-value in
the region around and larger than $\hat{p}_{(m)}$ are noises.

\begin{itemize}

\item[1.] Compute estimates of $p$-value. 
$$
  \hat{p}_i=(\text{false alarm rate of}~i\text{-th event}) \times \Delta t  
$$
where $i=1, ..., m$.
\item[2.] Let $\hat{p}_{(1)}\le\hat{p}_{(2)}\le\cdots\le\hat{p}_{(m)}$ be
  the ordered $p$-values.

\item[3.] Set
  $\hat{q}_{(m)}=n_{\rm obs}\hat{p}_{(m)}/m$. 
  
\item[4.] For $i=m-1,m-2,...,1$, compute
  \[
    \hat{q}_{(i)}=
    \min\left(\frac{n_{\rm obs}\hat{p}_{(i)}}{i},\hat{q}_{(i+1)}
    \right).
  \]
    
\item[5.] The estimated $q$-values for the $i$-th most significant event is
  $\hat{q}_{(i)}$.

\end{itemize}
\end{algorithm}

\subsection{1-OGC results}

In the 1-OGC catalog \cite{N18}, True Discovery Rate (TDR) and $P_{\rm astro}$ are given to evaluate the significance of events. 
A true discovery is the complement of the false discovery, FDR=1$-$TDR.  
Note however that the evaluation of TDR in \cite{N18} is a very conservative estimate. 
In \cite{N18}, an estimate of TDR is defined as 
\begin{equation}
\widehat{\text{TDR}}(\tilde{\rho}_c) = \frac{{\cal T}(\tilde{\rho}_c) }{{\cal T}(\tilde{\rho}_c) +{\cal F}(\tilde{\rho}_c) }, 
\label{eq:TDR}
\end{equation}
where ${\cal T}(\tilde{\rho}_c) $ is the rate that signals of astrophysical origin are observed with a ranking 
statistic $\geq \tilde{\rho}_c$, and ${\cal F}(\tilde{\rho}_c) $ is the FAR. 
In \cite{N18}, to estimate ${\cal T}(\tilde{\rho}_c)$, two significant events GW150914 and GW151229 are 
assumed to be real astrophysical signals, and ${\cal T}\sim 15{\rm yr}^{-1}$ is obtained. 
In order to take into account of the uncertainty in the estimate based on only two events, 
the Poisson  distribution is assumed for the observed number, and as a lower 95\% bound, 
${\cal T}\sim 2.7{\rm yr}^{-1}$ is obtained. 
In \cite{N18}, this value is used in (\ref{eq:TDR}) for all events other than GW150914 and GW151226. 

On the other hand, $P_{\rm astro}$ is the posterior probability given that a particular event has astrophysical origin.  In the 1-OGC catalog \cite{N18}, it is estimated as
\begin{equation}
P_{\rm astro}(\tilde{\rho}_c) = \frac{\Lambda_S P_S(\tilde{\rho}_c)}{\Lambda_N P_N(\tilde{\rho}_c) + \Lambda_S P_S(\tilde{\rho}_c)},
\label{eq:pastro}
\end{equation}
 where $P_S(\tilde{\rho}_c)$ and $P_N(\tilde{\rho}_c)$ are the probability densities of an event having ranking statistic $\tilde{\rho}_c$
given the event is signal or noise, respectively, and $\Lambda_S$ and $\Lambda_N$ are the rates of signal and noise events. \footnote{$P_{\rm astro}$ is also called {\it purity} in other field of physics \cite{ref:purity}.}
In order to estimate $\Lambda_S P_S(\tilde{\rho}_c)$, an analytic model of the signal distribution and a fixed conservative rate of mergers are used by assuming two events (GW150914 and GW151226) are astrophysical origin.
\footnote{Note that the method to estimate $P_{\rm astro}$ in \cite{N18} is different from that used in GWTC-1 catalog by LIGO-Virgo collaboration \cite{LV18} and in 2-OGC paper \cite{N20}.}

\begin{figure}[htp]
\begin{tabular}{cc}
\begin{minipage}{0.45\hsize}
\begin{center}
\includegraphics[width=\textwidth]{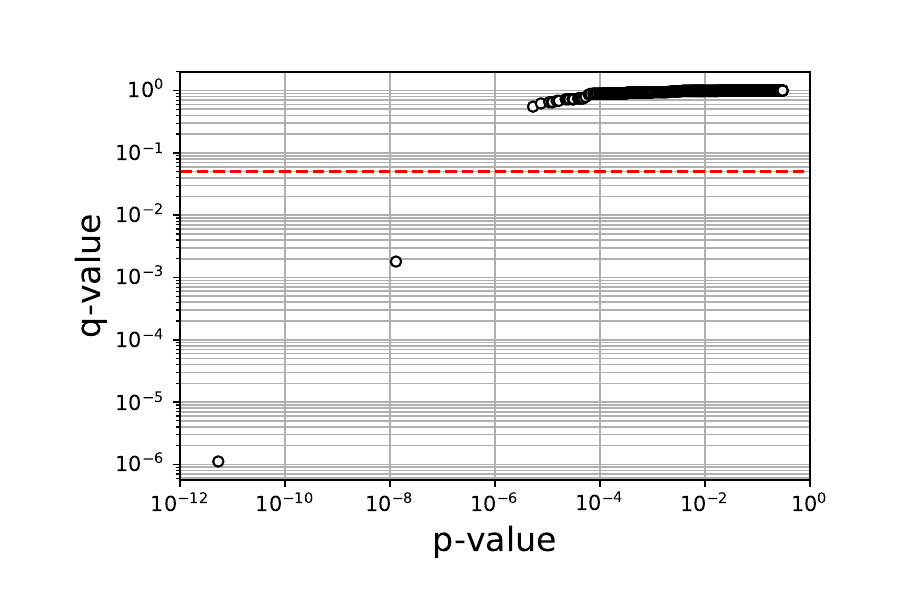}
\caption{
The $q$-values computed using Algorithm \ref{alg2}
from $p$-values of events in the \textit{complete} data set of 1-OGC.
The red dashed line indicates the $q$-value of 0.05.
}
\label{pq_all}
\end{center}
\end{minipage}
& %
\begin{minipage}{0.45\hsize}
\begin{center}
\includegraphics[width=\textwidth]{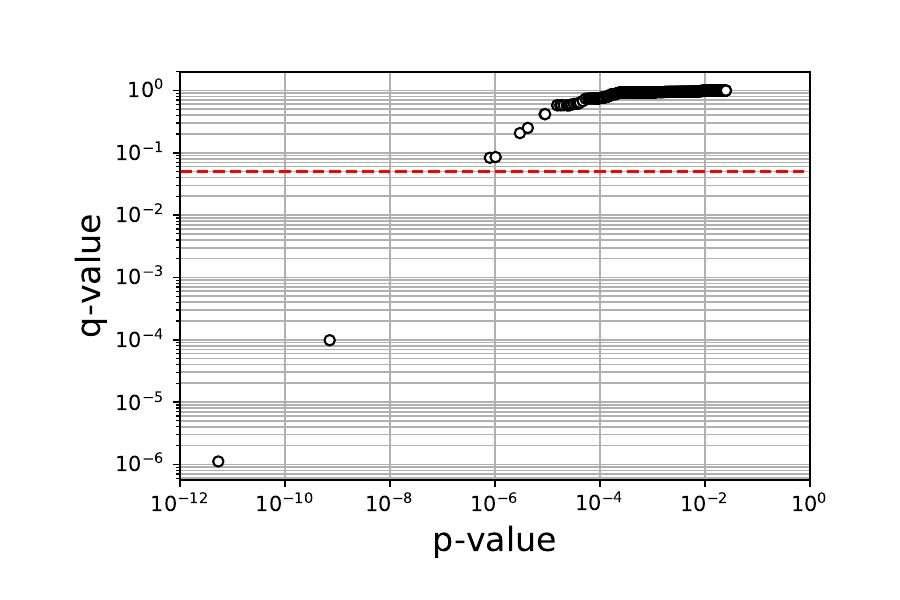}
\caption{
The $q$-values computed using Algorithm \ref{alg2}
from $p$-values of events in the \textit{bbh} data set of 1-OGC.
The red dashed line indicates the $q$-value of 0.05.
}
\label{pq_bbh}
\end{center}
\end{minipage}
\end{tabular}
\end{figure}

Figure \ref{pq_all}  shows
the $q$-value computed using Algorithm \ref{alg2} from $p$-value of events
in the \textit{complete} data set. 
Table \ref{table2} summarizes the results of the estimated $p$-value and $q$-value for 10 most significant events.  

Figure \ref{pq_bbh}  shows
the $q$-values computed using Algorithm \ref{alg2} from $p$-values of events
in the \textit{bbh} data set. 
Table \ref{table3} summarizes the results of estimated $p$-value and $q$-value for 10 most significant events. 
together with  the inverse of the false alarm rate, $1-\widehat{\text{TDR}}$ and $P_{\rm astro}$ given in the 1-OGC catalog. 
For the first two events, since only upper limit to the false alarm rate was evaluated in \cite{N18}, 
the estimated $p$-value of these events should be considered an upper limit to $p$-value. 
$1-\widehat{\text{TDR}}$ and $1-P_{\rm astro}$ are not given for the top two events in \cite{N18}, since these events are used 
to estimate $1-\widehat{\text{TDR}}$ and $P_{\rm astro}$ of other events. 

Following \cite{N18}, we discuss the significance of events with \textit{bbh} case. 
In Table \ref{table3}, if we call the events
whose $q$-value is smaller than $0.05$ significant, the top three events are
significant. The expected proportion of
false discoveries incurred in the three events is less than $0.05$. 
Since $q$-value of GW151012 (151012$+$09:54:43) is $9.83\times 10^{-5}$, 
this is significant enough as a true signal. 
In \cite{N18}, since $P_{\rm astro}$ for GW151012 is $9.76\times 10^{-1}$ which is larger than 0.5,
GW151012 is called significant. 
Thus, the results of $q$-value and $P_{\rm astro}$ are consistent for this events. 

In Table \ref{table3}, we find two marginally not significant events, 160103+05:48:36 and 151213+00:12:20
whose $q$-value are $8.31\times 10^{-2}$ and $8.53\times 10^{-2}$ respectively. 
On the other hand, $P_{\rm astro}$ for these events are small, $6.07\times 10^{-2}$ and $4.66\times 10^{-2}$ respectively.
So in \cite{N18}, these two events are called not significant. 
Although the conclusions are the same, the significance are slightly different between $q$-value and $P_{\rm astro}$ in \cite{N18}, and this difference might be interesting.
However, since these two events do not appear in the 2-OGC catalog in the next subsection, 
we don't investigate these events more. 

The value of $1-\widehat{\text{TDR}}$ is about 1 order of magnitude larger than $q$-value for all events. 
Since $\widehat{\text{TDR}}$ in \cite{N18} is a very conservative estimate, this difference is not surprising. 
Even in this case, $1-\widehat{\text{TDR}}$ for GW151012 is $8.29\times 10^{-4}$. Thus, this can be called significant. 
But $\widehat{\text{TDR}}$ for 160103+05:48:36 and 151213+00:12:20 is 0.483 and 0.545. 
Thus, these can not be called marginal events.  

In the LIGO-Virgo GWTC-1 catalog of gravitational-waves from compact binary mergers during O1 and O2 \cite{LV18},
a necessary condition that an event is
considered to be a gravitational wave signal is that the FAR of
the event is less than one per 30 days, which corresponds to the $p$-value of 
$10({\rm sec})/30({\rm days})=3.9\times 10^{-6}$.
By linearly fitting the data in Figs. \ref{pq_all} and \ref{pq_bbh}, 
we can evaluate that this $p$-value corresponds to 
the $q$-value of $0.411$ and $0.240$, respectively. 
The $q$-value of 0.05 corresponds to one per 271 days and one per 246 days of FAR, respectively. 
The $q$-value threshold of 0.05 is more stringent than the FAR of one per 30 days. 

When we compare $q$-value of same event,
the $q$-value in Table \ref{table3} is smaller than that in Table~\ref{table2}.
The reason for this difference is that events in the data set are computed 
from the different number of templates. The small number of templates 
decreases the false alarm rate and $p$-value. Accordingly, it produces different $q$-value.

\begin{table}[htp]
\begin{center}
\caption{Estimated $p$-values and $q$-values of the events 
of the \textit{complete} data set.
Events are sorted by false alarm rate and the 10 most significant events are shown.
The inverse false alarm rates ($\text{FAR}^{-1}$) are obtained from 1-OGC catalog.
$p$-values are computed by (\ref{p-value_est}).
$q$-values are computed by Algorithm \ref{alg2}.
}
\begin{tabular}{lcccc}
\hline
UTC time & $\text{FAR}^{-1}$ (year) & $p$-value  & $q$-value \\
\hline
150914+09:50:45  &  $>6.55\times 10^{4} $  &  $<4.84\times 10^{-12}$  &  $<1.11\times 10^{-6}$ \\
151226+03:38:53  &  $>5.91\times 10^{4} $  &  $<5.36\times 10^{-12}$  &  $<1.11\times 10^{-6}$ \\
151012+09:54:43  &  $2.44\times 10^{1} $  &  $1.30\times 10^{-8} $  &  $1.80\times 10^{-3}$ \\
151019+00:23:16  &  $5.96\times 10^{-2}$  &  $5.32\times 10^{-6} $  &  $5.52\times 10^{-1}$ \\
150928+10:49:00  &  $4.24\times 10^{-2}$  &  $7.48\times 10^{-6} $  &  $6.22\times 10^{-1}$ \\
151218+18:30:58  &  $2.93\times 10^{-2}$  &  $1.08\times 10^{-5} $  &  $6.51\times 10^{-1}$ \\
160103+05:48:36  &  $2.63\times 10^{-2}$  &  $1.21\times 10^{-5} $  &  $6.51\times 10^{-1}$ \\
151202+01:18:13  &  $2.53\times 10^{-2}$  &  $1.25\times 10^{-5} $  &  $6.51\times 10^{-1}$ \\
160104+03:51:51  &  $2.12\times 10^{-2}$  &  $1.49\times 10^{-5} $  &  $6.84\times 10^{-1}$ \\
151213+00:12:20  &  $1.93\times 10^{-2}$  &  $1.64\times 10^{-5} $  &  $6.84\times 10^{-1}$ \\
\hline
\end{tabular}
\label{table2}
\end{center}
\end{table}

\begin{table}[htp]
\begin{center}
\caption{
The same as Table \ref{table2}, but obtained from \textit{bbh} data set.
$\text{FAR}$, $1-\widehat{\text{TDR}}$ and $1-P_{\rm astro}$ are obtained from 1-OGC catalog.
}
\begin{tabular}{lc|cc|cc}
\hline
UTC time  & $\text{FAR}^{-1}$ (year) &  $p$-value & $q$-value &  $1-\widehat{\text{TDR}}$  & $P_{\rm astro}$ \\
\hline
150914+09:50:45  &  $>6.55\times 10^{4} $  &  $<4.84\times 10^{-12}$ &  $<1.11\times 10^{-6}$  &  $\cdots$  &  $\cdots$ \\
151226+03:38:53  &  $>5.91\times 10^{4} $  &  $<5.36\times 10^{-12}$ &  $<1.11\times 10^{-6}$  &  $\cdots$  &  $\cdots$ \\
151012+09:54:43  &  $4.46\times 10^{2} $   &  $7.10\times 10^{-10}$  &  $9.83\times 10^{-5}$   &  $8.29\times 10^{-4}$  &  $9.76\times 10^{-1}$ \\
\hdashline
160103+05:48:36  &  $3.96\times 10^{-1}$   &  $8.00\times 10^{-7}$   &  $8.31\times 10^{-2}$   &  $4.83\times 10^{-1}$  &  $6.07\times 10^{-2}$ \\
151213+00:12:20  &  $3.09\times 10^{-1}$   &  $1.03\times 10^{-6}$   &  $8.53\times 10^{-2}$   &  $5.45\times 10^{-1}$  &  $4.66\times 10^{-2}$ \\
\hdashline
151216+18:49:30  &  $1.06\times 10^{-1}$   &  $2.98\times 10^{-6}$   &  $2.07\times 10^{-1}$   &  $7.77\times 10^{-1}$  &  $1.72\times 10^{-2}$ \\
151222+05:28:26  &  $7.51\times 10^{-2}$   &  $4.22\times 10^{-6}$   &  $2.50\times 10^{-1}$   &  $8.31\times 10^{-1}$  &  $1.20\times 10^{-2}$ \\
151217+03:47:49  &  $3.59\times 10^{-2}$   &  $8.82\times 10^{-6}$   &  $4.17\times 10^{-1}$   &  $9.12\times 10^{-1}$  &  $5.99\times 10^{-3}$ \\
151009+05:06:12  &  $3.51\times 10^{-2}$   &  $9.02\times 10^{-6}$   &  $4.17\times 10^{-1}$   &  $9.13\times 10^{-1}$  &  $5.20\times 10^{-3}$ \\
151220+07:45:36  &  $2.07\times 10^{-2}$   &  $1.53\times 10^{-5}$   &  $5.78\times 10^{-1}$   &  $9.47\times 10^{-1}$  &  $3.20\times 10^{-3}$ \\
\hline
\end{tabular}
\label{table3}
\end{center}
\end{table}

\subsection{2-OGC results}

\begin{figure}[htp]
\begin{tabular}{cc}
\begin{minipage}{0.45\hsize}
\begin{center}
\includegraphics[width=\textwidth]{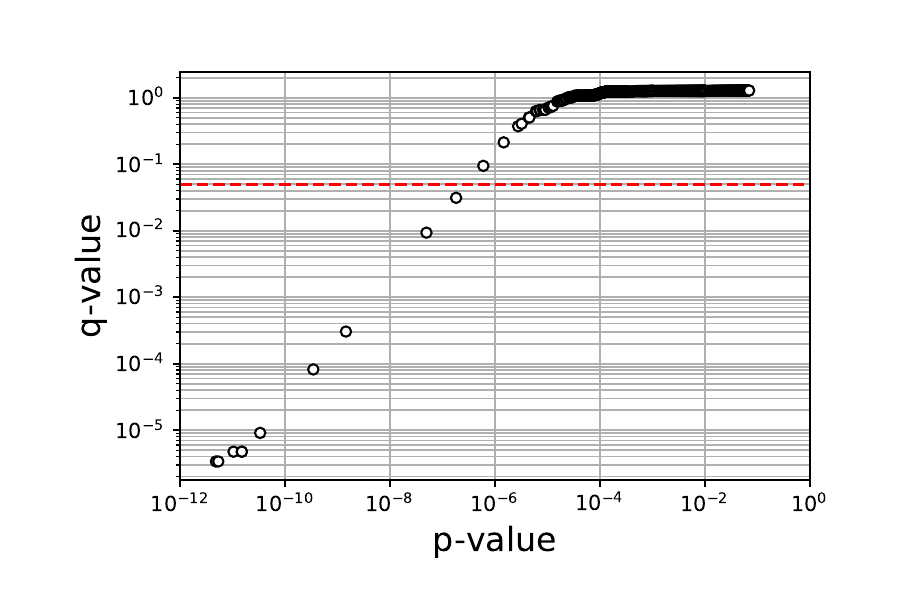}
\caption{
The $q$-values computed using Algorithm \ref{alg2}
from $p$-values of events in the \textit{complete} data set of 2-OGC.
The red dashed line indicates the $q$-value of 0.05.
}
\label{pq_all_2}
\end{center}
\end{minipage}
& %
\begin{minipage}{0.45\hsize}
\begin{center}
\includegraphics[width=\textwidth]{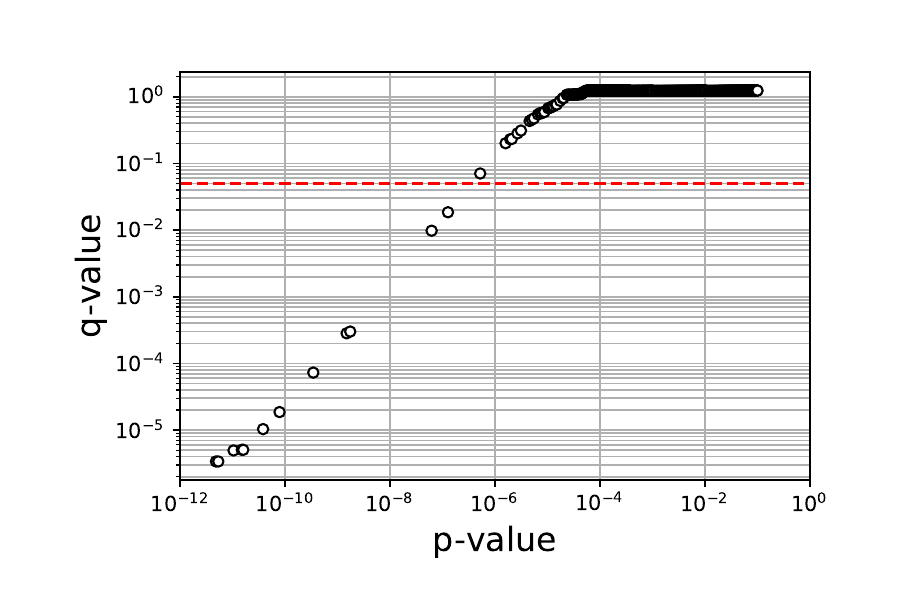}
\caption{
The $q$-values computed using Algorithm \ref{alg2}
from $p$-values of events in the \textit{bbh} data set of 2-OGC.
The red dashed line indicates the $q$-value of 0.05.
}
\label{pq_bbh_2}
\end{center}
\end{minipage}
\end{tabular}
\end{figure}

Figure \ref{pq_all_2}  shows the $q$-values as a function of $p$-values 
in the \textit{complete} data set. 
Table \ref{tab:2-OGC-all} summarizes the results of the estimated $q$-values of the events 
for 30 most significant events.

Figure \ref{pq_bbh_2}  shows the $q$-values as a function of $p$-values 
in the \textit{bbh} data set. 
Table \ref{tab:2-OGC-bbh} summarizes the results of estimated $q$-values for top 30 events. 
$P_{\rm astro}$ computed in the 2-OGC paper \cite{N20} is also shown in this table. \footnote{The method to estimate $P_{\rm astro}$ in \cite{N20} is based on a mixture model developed in Farr et al. \cite{FGMC15} and employed in GWTC-1 catalog by LIGO-Virgo collaboration \cite{LV18}.}

We discuss the significance of events for \textit{bbh} case. 
In table \ref{tab:2-OGC-bbh}, if we call the events
whose $q$-value is smaller than $0.05$ significant, the top 13 events are called significant. 
In \cite{N20}, these 13 events are called significant since $P_{\rm astro}$ is larger than 0.5. 
Thus, the results of $q$-value and $P_{\rm astro}$ are consistent each other. 
On the other hand, we obtain a different result for 151205+19:55:25. 
The $q$-value of this event is 0.07, while $P_{\rm astro}$ is 0.525.
Thus, this is definitely a marginal event. 
If we call events with $q$-value less than 0.05 significant, 
this event can not be called significant. 
On the other hand, in \cite{N20}, this event is called significant, since $P_{\rm astro}$ is larger than 0.5,
and is identified as a new marginal binary black hole merger, GW151205.

Finally, we investigate correspondence between $q$-value and FAR. 
By linearly fitting the data in Figs. \ref{pq_all_2} and \ref{pq_bbh_2}, 
we can evaluate that $p$-value of $10({\rm sec})/30({\rm days})=3.9\times 10^{-6}$ 
corresponds to the $q$-value of $0.463$ and $0.377$, respectively.  
The $q$-value of 0.05 corresponds to one per 384 days and one per 319 days of FAR, respectively. 
Thus, as in the case of 1-OGC, 
the $q$-value threshold of 0.05 is more stringent than the FAR of one per 30 days.

\begin{table}[htp]
\begin{center}
\caption{Estimated $p$-values and $q$-values of the events 
of the \textit{complete} data set of 2-OGC. 
Events are sorted by false alarm rate and the top 30 events are shown.
The inverse false alarm rates ($\text{FAR}^{-1}$) are obtained from the data set.
$p$-values are computed by (\ref{p-value_est}).
$q$-values are computed by Algorithm \ref{alg2}.
}
\begin{tabular}{lc|cc}
\hline
UTC time & $\text{FAR}^{-1}$ (year) &  $p$-value  & $q$-value \\
\hline
170104+10:11:58  &  $>7.27\times 10^{4} $  &  $<4.36\times 10^{-12}$  &  $<3.38\times 10^{-6}$ \\
150914+09:50:45  &  $>6.55\times 10^{4} $  &  $<4.84\times 10^{-12}$  &  $<3.38\times 10^{-6}$ \\ 
151226+03:38:53  &  $>5.91\times 10^{4} $  &  $<5.36\times 10^{-12}$  &  $<3.38\times 10^{-6}$ \\
170823+13:13:58  &  $>3.03\times 10^{4} $  &  $<1.05\times 10^{-11}$  &  $<4.75\times 10^{-6}$ \\
170817+12:41:04  &  $>2.11\times 10^{4} $  &  $<1.51\times 10^{-11}$  &  $<4.75\times 10^{-6}$ \\
170814+10:30:43  &  $>2.11\times 10^{4} $  &  $<1.51\times 10^{-11}$  &  $<4.75\times 10^{-6}$ \\
170809+08:28:21  &  $9.42\times 10^{3} $  &  $3.36\times 10^{-11}$  &  $9.10\times 10^{-6}$ \\
170608+02:01:16  &  $>9.15\times 10^{2} $  &  $<3.46\times 10^{-10}$  &  $<8.19\times 10^{-5}$ \\
151012+09:54:43  &  $2.19\times 10^{2} $  &  $1.45\times 10^{-9} $  &  $3.04\times 10^{-4}$ \\
170729+18:56:29  &  $6.41              $          &  $4.95\times 10^{-8} $  &  $9.36\times 10^{-3}$ \\
170121+21:25:36  &  $1.74              $          &  $1.82\times 10^{-7} $  &  $3.13\times 10^{-2}$ \\
170727+01:04:30  &  $5.28\times 10^{-1}$  &  $6.00\times 10^{-7} $  &  $9.46\times 10^{-2}$ \\
170818+02:25:09  &  $2.16\times 10^{-1}$  &  $1.46\times 10^{-6} $  &  $2.13\times 10^{-1}$ \\
170722+08:45:14  &  $1.15\times 10^{-1}$  &  $2.76\times 10^{-6} $  &  $3.73\times 10^{-1}$ \\
170321+03:13:21  &  $9.84\times 10^{-2}$  &  $3.22\times 10^{-6} $  &  $4.06\times 10^{-1}$ \\
170310+09:30:52  &  $7.17\times 10^{-2}$  &  $4.42\times 10^{-6} $  &  $5.04\times 10^{-1}$ \\
170809+03:55:52  &  $7.00\times 10^{-2}$  &  $4.53\times 10^{-6} $  &  $5.04\times 10^{-1}$ \\
170819+07:30:53  &  $5.26\times 10^{-2}$  &  $6.02\times 10^{-6} $  &  $6.29\times 10^{-1}$ \\
170618+20:00:39  &  $5.02\times 10^{-2}$  &  $6.32\times 10^{-6} $  &  $6.29\times 10^{-1}$ \\
170416+18:38:48  &  $4.47\times 10^{-2}$  &  $7.09\times 10^{-6} $  &  $6.53\times 10^{-1}$ \\
170331+07:08:18  &  $4.37\times 10^{-2}$  &  $7.25\times 10^{-6} $  &  $6.53\times 10^{-1}$ \\
151216+18:49:30  &  $3.88\times 10^{-2}$  &  $8.17\times 10^{-6} $  &  $6.59\times 10^{-1}$ \\
170306+04:45:50  &  $3.64\times 10^{-2}$  &  $8.71\times 10^{-6} $  &  $6.59\times 10^{-1}$ \\
151227+16:52:22  &  $3.62\times 10^{-2}$  &  $8.76\times 10^{-6} $  &  $6.59\times 10^{-1}$ \\
170126+23:56:22  &  $3.54\times 10^{-2}$  &  $8.95\times 10^{-6} $  &  $6.59\times 10^{-1}$ \\
151202+01:18:13  &  $3.50\times 10^{-2}$  &  $9.06\times 10^{-6} $  &  $6.59\times 10^{-1}$ \\
170208+20:23:00  &  $3.02\times 10^{-2}$  &  $1.05\times 10^{-5} $  &  $7.11\times 10^{-1}$ \\
170327+17:07:35  &  $3.01\times 10^{-2}$  &  $1.05\times 10^{-5} $  &  $7.11\times 10^{-1}$ \\
170823+13:40:55  &  $2.75\times 10^{-2}$  &  $1.15\times 10^{-5} $  &  $7.26\times 10^{-1}$ \\
150928+10:49:00  &  $2.75\times 10^{-2}$  &  $1.15\times 10^{-5} $  &  $7.26\times 10^{-1}$ \\
\hline
\end{tabular}
\label{tab:2-OGC-all}
\end{center}
\end{table}

\begin{table}[htp]
\begin{center}
\caption{Estimated $p$-values and $q$-values of the events 
of the \textit{bbh} data set of 2-OGC. 
Events are sorted by the inverse false alarm rate ($\text{FAR}^{-1}$) and the top 30 events are shown.
The inverse false alarm rates are obtained from the data set.
$p$-values are computed by (\ref{p-value_est}).
$q$-values are computed by Algorithm \ref{alg2}.
}
\begin{tabular}{lc|cc|c}
\hline
UTC time & $\text{FAR}^{-1}$ (year) &  $p$-value  & $q$-value & $P_{\rm astro}$\\
\hline
170104+10:11:58  &  $>7.27\times 10^{4}$   &  $<4.36\times 10^{-12}$  &  $<3.38\times 10^{-6}$ &  $>0.999$\\
150914+09:50:45  &  $>6.55\times 10^{4}$   &  $<4.84\times 10^{-12}$  &  $<3.38\times 10^{-6}$ &  $>0.999$\\
151226+03:38:53  &  $>5.91\times 10^{4}$   &  $<5.36\times 10^{-12}$  &  $<3.38\times 10^{-6}$ &  $>0.999$\\
170823+13:13:58  &  $>3.03\times 10^{4}$   &  $<1.05\times 10^{-11}$  &  $<4.95\times 10^{-6}$ &  $>0.999$\\
170814+10:30:43  &  $>2.11\times 10^{4}$   &  $<1.51\times 10^{-11}$  &  $<5.06\times 10^{-6}$ &  $>0.999$\\
151012+09:54:43  &  $>1.98\times 10^{4}$   &  $<1.60\times 10^{-11}$  &  $<5.06\times 10^{-6}$ &  $>0.999$\\
170809+08:28:21  &  $8.28\times 10^{3}$    &  $3.83\times 10^{-11}$   &  $1.03\times 10^{-5}$  &  $>0.999$\\
170729+18:56:29  &  $4.02\times 10^{3}$    &  $7.88\times 10^{-11}$   &  $1.86\times 10^{-5}$  &  $>0.999$\\
170608+02:01:16  &  $>9.15\times 10^{2}$   &  $<3.46\times 10^{-10}$  &  $<7.28\times 10^{-5}$ &  $>0.999$\\
170121+21:25:36  &  $2.12\times 10^{2}$    &  $1.49\times 10^{-9}$    &  $2.83\times 10^{-4}$  &  $>0.999$\\
170727+01:04:30  &  $1.81\times 10^{2}$    &  $1.75\times 10^{-9}$    &  $3.01\times 10^{-4}$  &  $9.94\times 10^{-1}$\\
170818+02:25:09  &  $5.11\times 10^{0}$    &  $6.21\times 10^{-8}$    &  $9.79\times 10^{-3}$  &  $>0.999$\\
170304+16:37:53  &  $2.49\times 10^{0}$    &  $1.27\times 10^{-7}$    &  $1.85\times 10^{-2}$  &  $6.97\times 10^{-1}$\\
\hdashline
151205+19:55:25  &  $6.07\times 10^{-1}$   &  $5.22\times 10^{-7}$    &  $7.06\times 10^{-2}$  &  $5.25\times 10^{-1}$\\
\hdashline
170425+05:53:34  &  $1.99\times 10^{-1}$   &  $1.59\times 10^{-6}$    &  $2.01\times 10^{-1}$  &  $2.05\times 10^{-1}$\\
170201+11:03:12  &  $1.63\times 10^{-1}$   &  $1.95\times 10^{-6}$    &  $2.30\times 10^{-1}$  &  $2.39\times 10^{-1}$\\
151217+03:47:49  &  $1.52\times 10^{-1}$   &  $2.09\times 10^{-6}$    &  $2.33\times 10^{-1}$  &  $2.57\times 10^{-1}$\\
151011+19:27:49  &  $1.18\times 10^{-1}$   &  $2.69\times 10^{-6}$    &  $2.82\times 10^{-1}$  &  $7.95\times 10^{-2}$\\
151216+09:24:16  &  $1.01\times 10^{-1}$   &  $3.12\times 10^{-6}$    &  $3.11\times 10^{-1}$  &  $1.81\times 10^{-1}$\\
170403+23:06:11  &  $6.93\times 10^{-2}$   &  $4.57\times 10^{-6}$    &  $4.33\times 10^{-1}$  &  $3.26\times 10^{-2}$\\
170202+13:56:57  &  $6.29\times 10^{-2}$   &  $5.04\times 10^{-6}$    &  $4.54\times 10^{-1}$  &  $1.28\times 10^{-1}$\\
170629+04:13:55  &  $5.76\times 10^{-2}$   &  $5.50\times 10^{-6}$    &  $4.73\times 10^{-1}$  &  $1.90\times 10^{-2}$\\
170220+11:36:24  &  $4.81\times 10^{-2}$   &  $6.59\times 10^{-6}$    &  $5.42\times 10^{-1}$  &  $1.03\times 10^{-1}$\\
170721+05:55:13  &  $4.40\times 10^{-2}$   &  $7.20\times 10^{-6}$    &  $5.67\times 10^{-1}$  &  $5.99\times 10^{-2}$\\
170123+20:16:42  &  $4.07\times 10^{-2}$   &  $7.78\times 10^{-6}$    &  $5.70\times 10^{-1}$  &  $8.41\times 10^{-2}$\\
170801+23:28:19  &  $4.05\times 10^{-2}$   &  $7.83\times 10^{-6}$    &  $5.70\times 10^{-1}$  &  $\cdots$\\
170818+09:34:45  &  $3.69\times 10^{-2}$   &  $8.58\times 10^{-6}$    &  $5.90\times 10^{-1}$  &  $\cdots$\\
170620+01:14:02  &  $3.63\times 10^{-2}$   &  $8.73\times 10^{-6}$    &  $5.90\times 10^{-1}$  &  $1.51\times 10^{-2}$\\
151216+18:49:30  &  $3.07\times 10^{-2}$   &  $1.03\times 10^{-5}$    &  $6.73\times 10^{-1}$  &  $6.93\times 10^{-2}$\\
170104+21:58:40  &  $2.87\times 10^{-2}$   &  $1.10\times 10^{-5}$    &  $6.83\times 10^{-1}$  &  $1.19\times 10^{-1}$\\
\hline
\end{tabular}
\label{tab:2-OGC-bbh}
\end{center}
\end{table}

\section{Summary and Discussion}
\label{sec5}

In this paper, we presented
a consistent procedure to assess the significance of each event.
We proposed an estimator of the $p$-values (\ref{p-value_est})
of a particular event in the statistical hypothesis testing
by using the empirical distribution of detection statistic without any assumption on the background distribution.
Generally, $p$-value should follows a uniform distribution if all events are originated from noise. 
The $p$-value defined in (\ref{p-value_est}) has this property. 
On the other hand, the $p_{\rm conv}$ defined in  (\ref{p-value_conv}) does not have this property in general. 
We thus believe that the $p$-value in (\ref{p-value_est}) is more useful to assess the significance of each event
than $p_{\rm conv}$ in (\ref{p-value_conv}). 
Moreover, we proposed a consistent procedure to evaluate $q$-value which is a measure of FDR. 
In this procedure, we use a property that $p$-values follow the uniform distribution  under the null hypothesis, and
we don't need any assumptions on the distribution of signals. 
We apply this procedure to 1-OGC and 2-OGC catalog data \cite{N18}\cite{N20}. 
There is already a procedure which was introduced to evaluate $q$-value in the literature \cite{ST03}. 
However, since not all events in the analysis are available in the catalogs, 
we proposed a new procedure to evaluate $q$-value which is a modified version of the original one. 

The results are shown in Tables \ref{table2}, \ref{table3}, \ref{tab:2-OGC-all} and \ref{tab:2-OGC-bbh}. 
For \textit{bbh} case of 1-OGC, if we call events with $q$-value less than 0.05 significant, 
we have 3 significant events, GW150914, GW151226 and GW151012. 
This is fully consistent with the conclusion of 1-OGC paper \cite{N18}. 
We also found 2 marginally not significant events, 160103+05:48:36 and 151213+00:12:20, 
whose $q$-value are $8.31\times 10^{-2}$ and $8.53\times 10^{-2}$, respectively. 
Since $P_{\rm astro}$ for these events are $6.07\times 10^{-2}$ and $4.66\times 10^{-2}$, these are not 
identified as marginal events in \cite{N18}.  

For \textit{bbh} case of 2-OGC, we have 13 significant events. All of them are also identified significant 
based on $P_{\rm astro}$ in \cite{N20}. 
There is one marginal event, 151205+19:55:25. 
The $q$-value of this event is 0.07 but the $P_{\rm astro}$ computed in \cite{N20} is 0.525. 
Thus, $q$-value suggests this is marginally not significant, while $P_{\rm astro}$ suggests this is 
marginally significant. 
It is not easy to conclude whether this signal is from astrophysical origin or not only from these results. 

The method for estimating $q$-value presented in this paper is very simple because we don't need any assumptions on the distribution of noise and signal.
Note that $q$-value and $P_{\rm astro}$ are based on fundamentally distinct statistical disciplines. The $q$-value is a frequentist  measure, which is devised to estimate FDR of events over some threshold of significance without any assumptions on signals. In contrast, $P_{\rm astro}$
is a Bayesian measure, which is devised to estimate the posterior
probability of astrophysical origin of a particular event relying on prior assumptions on signals.
Nevertheless, from the results discussed above, we found that both approaches provide almost the same
conclusion. The coincidence is not at all trivial. The coincidence
would suggest that the prior assumptions on signals used in the
computation of $P_{\rm astro}$ are close to the reality.
It should be useful to estimate $q$-value as well as $P_{\rm astro}$ in the gravitational wave searches. 
This should be true especially for marginal events like 151205+19:55:25 in this paper. We can obtain additional information on the significance of an event from different criterion. 

We also note that the procedure  for estimating $q$-value presented in this paper 
can be applicable to other searches for gravitational waves. 
Our procedure for estimating $q$-value is not restricted to the specific
searches for the gravitational waves
whose true background distribution of detection statistic is difficult to know,
because our procedure is based on the 
empirical distribution, which is always available by time-shifting of
time-series data of different detectors.

\section*{Acknowledgment}
H.Y. and H.T. would like to thank Jishnu Suresh for fruitful discussion.
We thank the authors of 1-OGC \cite{N18} and 2-OGC \cite{N20} for making the data set of the catalogs public. 
This work was supported by MEXT, JSPS Leading-edge Research Infrastructure Program, JSPS Grant-in-Aid for Specially Promoted Research 26000005, JSPS Grant-in-Aid for Scientific Research on Innovative Areas 2905: JP17H06358, JP17H06361, JP16H02183 and JP17H06364, JSPS Core-to-Core Program A. Advanced Research Networks, JSPS Grant-in-Aid for Scientific Research (S) 17H06133 and 15H00787, the joint research program of the Institute for Cosmic Ray Research, 
the cooperative research program of the Institute of Statistical Mathematics, National Research Foundation (NRF) and Computing Infrastructure Project of KISTI-GSDC in Korea, Academia Sinica (AS), AS Grid Center (ASGC) and the Ministry of Science and Technology (MoST) in Taiwan under grants including AS-CDA-105-M06, Advanced Technology Center (ATC) of NAOJ, Mechanical Engineering Center of KEK, the LIGO project, and the Virgo project.

\appendix

\section{Derivation and meaning of $p_{\rm conv}$}
As in various scientific research fields \cite{ASA}, there might be some 
confusion in use of $p$-value in the gravitational wave community.
In the recent American statistical association statement on $p$-value
\cite{ASA}, the first principle is ``$P$-values can indicate how incompatible 
the data are with a specified statistical model". Therefore, if we are
saying about a $p$-value, we always have to make clear what statistical model 
we are talking about.
In this appendix, we discuss derivation and meaning of the 
conventional $p$-value $p_\text{conv}$ defined by
(\ref{p-value_conv}), which is the probability of observing one or more noise events as strong as a signal 
whose detection statistic is $\rho$ under the noise model. 
In the analysis paper of the event GW150914 \cite{A16}, Abbott et al. called 
$p_\text{conv}$ a $p$-value, however, in the text we have not called it 
$p$-value to avoid a possible confusion with the $p$-value 
defined by (\ref{p-value}).

Let us see more details of  
the probability (\ref{p-value_conv}) which
was proposed by Usman et al. in Appendix of
\cite{U16}. The total number of noise events in the observed data,  $N$, is modeled parametrically
with the Poisson process of mean $\mu$:
\begin{equation}
  \mathbb{P}(N=n)=\frac{\mu^n}{n!}e^{-\mu}, \qquad n\in\{0,1,2,...\},
  \label{poisson}
\end{equation}
where $\mu=\mu(\rho)$.
The slight difference between the expression of $\mu(\rho)$
in (\ref{p-value_conv})
and the expression $(1+n_{\rm bg}(\rho)t_{\rm obs})/t_{\rm bg}$ in 
Eq. 17
of \cite{U16} (the unity in the numerator) comes from the fact that the model
used by Usman et al. \cite{U16} involves observed events. In contrast,
(\ref{p-value_conv}) is based only on the noise events in simulated background
data, because the authors of the present paper believe that the noise model is
better to be constructed by noise events only. In addition, Usman et al
\cite{U16} considered a randomness in the number of candidate events and then
marginalized them out. However, these steps have no influence on the final expression if
$n_{\rm bg}(\rho)\ll n_{\rm bg}(0)$ (compare Equations A.4 and A.12 in
\cite{U16}). Then, the probability of observing one or more noise events as
strong as a signal whose detection statistic is $\rho$ under the noise model
during the observation time,
$\mathbb{P}(N\ge 1)$, is given by (\ref{p-value_conv}). 
In the same manner, if we consider the probability of observing $n_0$ or
more noise events as strong as a signal whose detection statistic is $\rho$
under the noise model during the observation time, the $p$-value is
\[
p_\text{conv}(\rho;n_0) := \mathbb{P}(N\ge n_0)
=\sum_{n\ge n_0}\frac{\mu^n}{n!}e^{-\mu}.
\label{p-value_conv_mod}
\]

\section{Discussion on $\hat{\pi}_0$}
\label{appendixpi0}

\begin{figure}[tph]
\begin{tabular}{cc}
\begin{minipage}[t]{0.45\hsize}
\begin{center}
\includegraphics[width=\textwidth]{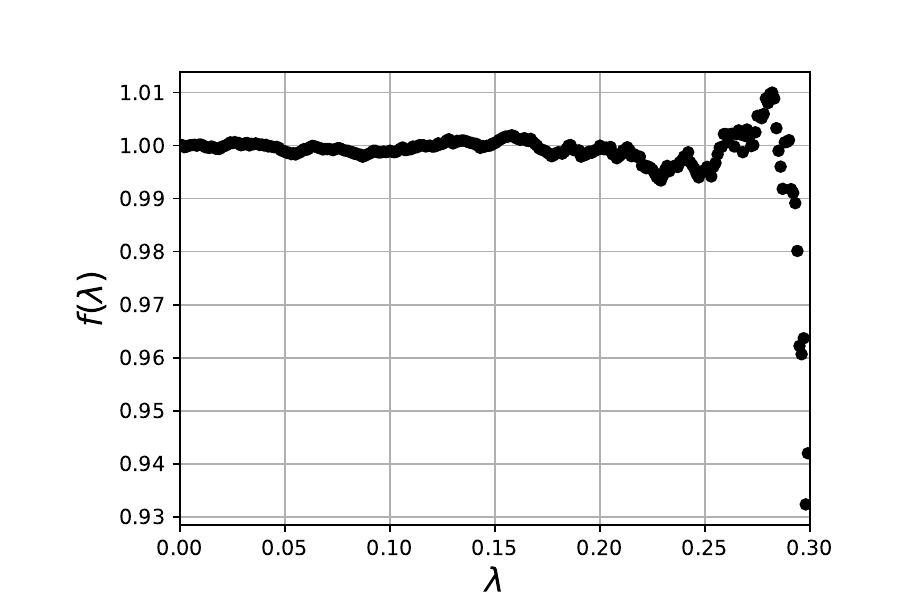}
\caption{
Plot of $f(\lambda)$ defined in \eqref{flambda} 
for the \textit{complete} data set of 1-OGC. 
}
\label{plot_f_lambda_1_all} 
\end{center}
\end{minipage}
& %
\begin{minipage}[t]{0.45\hsize}
\begin{center}
\includegraphics[width=\textwidth]{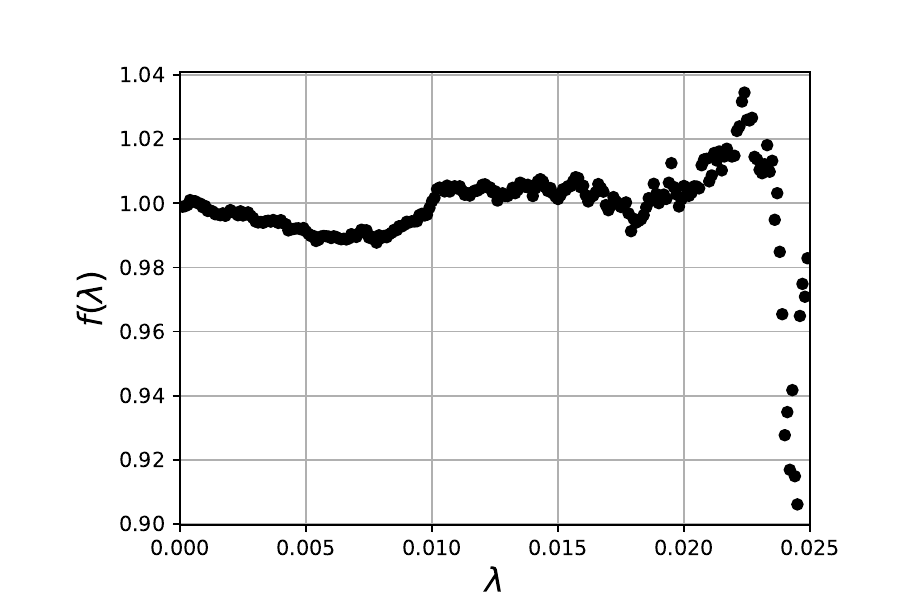}
\caption{
Plot of $f(\lambda)$ defined in \eqref{flambda}
for the \textit{bbh} data set of 1-OGC. 
}
\label{plot_f_lambda_1_bbh} 
\end{center}
\end{minipage}
\end{tabular}
\end{figure}

\begin{figure}[htp]
\begin{tabular}{cc}
\begin{minipage}[t]{0.45\hsize}
\begin{center}
\includegraphics[width=\textwidth]{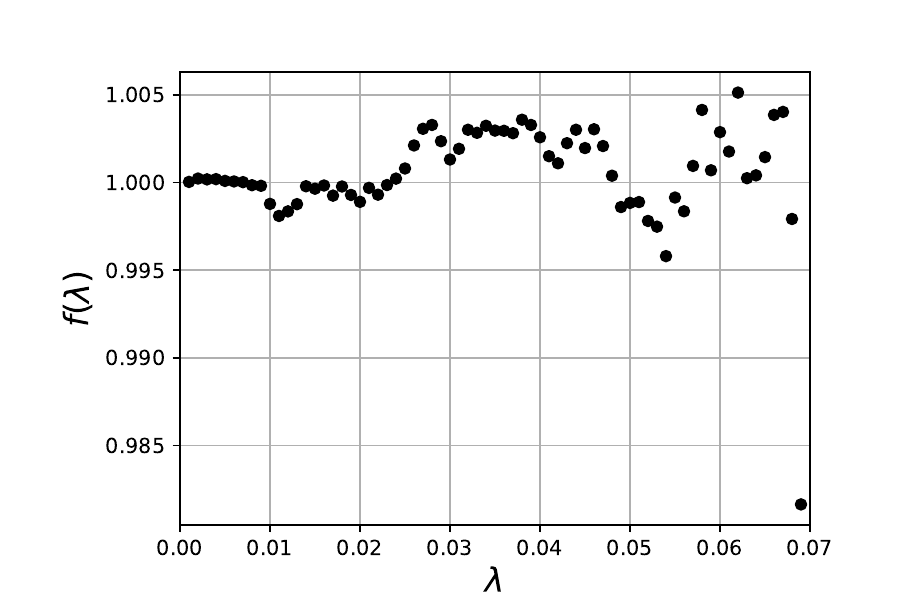}
\caption{
Plot of $f(\lambda)$ defined in \eqref{flambda} 
for the \textit{complete} data set of 2-OGC. 
}
\label{plot_f_lambda_2_all} 
\end{center}
\end{minipage}
& %
\begin{minipage}[t]{0.45\hsize}
\begin{center}
\includegraphics[width=\textwidth]{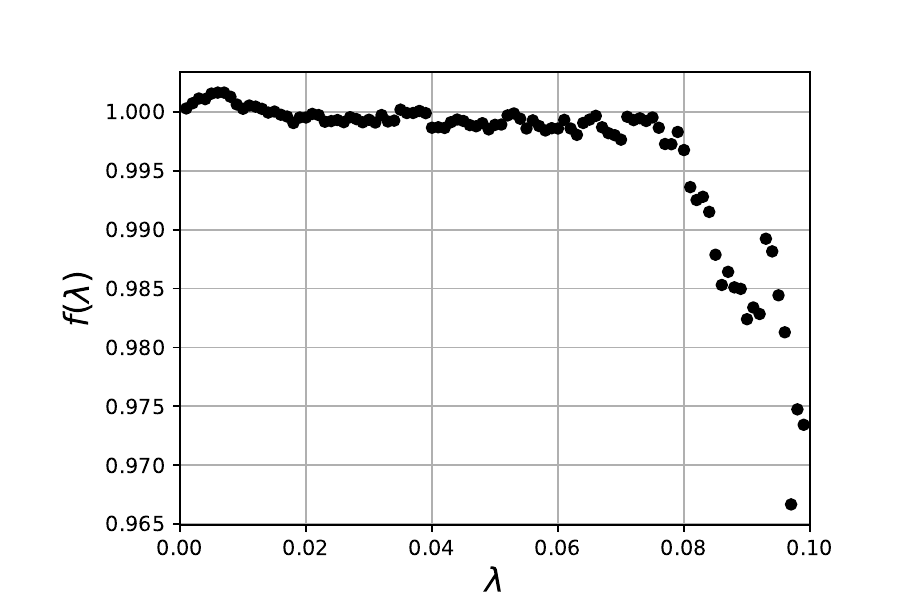}
\caption{
Plot of $f(\lambda)$ defined in \eqref{flambda}
for the \textit{bbh} data set of 2-OGC. 
}
\label{plot_f_lambda_2_bbh} 
\end{center}
\end{minipage}
\end{tabular}
\end{figure}

In this Appendix, we show that $\hat{\pi}_0$ in Eq.(\ref{FDRt_est}) can be approximated 
to be $\hat{\pi}_0\simeq 1$. 
$\hat{\pi}_0$ is an estimator of $\pi_0=n_0/n_{\rm obs}$
which indicates the overall proportion of noise events in the data. 
Setting $\hat{\pi}_0=1$ is reasonable when very few events are expected to be signal, such 
as the gravitational wave search.
In fact, Benjamini and Hochberg's proposal \cite{BH95} was setting 
$\hat{\pi}_0=1$. On the other hands, 
for data in which some portion of events are expected to be signal,
such as in genomewide studies, Storey and Tibshirani \cite{ST03} proposed
$\hat{\pi}_0=\hat{f}(1)$,
where $\hat{f}(\lambda)${, $\lambda\in(0,1)$} is an estimate of $f(\lambda)$ disccused in \eqref{flambda}.

We consider a list of $p$-values which contains $m$ $p$-values less than a certain value, and set $n_{\rm obs}=m$. We assume that the maximum $p$-value in this list is $\hat{p}_{(m)}$. In this case, $n_0$ is the number of noise whose $p$-values are between $0$ and $\hat{p}_{(m)}$. We consider a function
\begin{equation}
n(\lambda) = \frac{\#\{\hat{p}_i > \lambda; i\in\{1,\ldots,m\} \} }{1-\lambda/\hat{p}_{(m)}}
\end{equation}
where $0 < \lambda < \hat{p}_{(m)}$.
If all $p$-values larger than $\lambda_0$ are noise, $\mathbb{E}(n(\lambda))=n_0$ for $\lambda>\lambda_0$, since $p$-values follow uniform distribution.

As an estimator of $\hat{\pi}_0$, let us consider a function
\begin{equation}
f(\lambda)=\frac{n(\lambda)}{m}=\frac{\#\{\hat{p}_i > \lambda; i\in\{1,\ldots,m\} \}}{m(1-\lambda/\hat{p}_{(m)}) } 
\label{flambda}
\end{equation}
where $0 < \lambda < \hat{p}_{(m)}$.
If all $p$-values larger than $\lambda_0$ are noise, 
$\mathbb{E}(f(\lambda))=\pi_0$ for $\lambda>\lambda_0$.
In particular, if all $p$-values are noise, $\mathbb{E}(f(\lambda))=1$ for
$\lambda\in(0,\hat{p}_{(m)})$.

Figure \ref{plot_f_lambda_1_all} is the plot of $f(\lambda)$ for the \textit{complete} data set of 1-OGC. 
In this plot, we use $m=124,524$ events whose $p$-value is less than 0.3. 
We can see that $f(\lambda)$ in Eq. (\ref{flambda}) is almost unity for $0 < \lambda < 0.25$.
We have $0.99 < f(\lambda) <1.01$ in this region.  
This means that almost all $p$-values are noise 
except for a very few $p$-values around zero. 
Larger scatter in $0.25 < \lambda < 0.3$ is due to the statistical fluctuation caused by the 
smaller number in the numerator of Eq. (\ref{flambda}).
Since we are mainly interested in the events with small $p$-value less than $10^{-2}$, 
we set $\hat{\pi}_0=1$. 

The situation is similar to the \textit{bbh} case. 
Figure \ref{plot_f_lambda_1_bbh} is the plot of $f(\lambda)$ for $0 < \lambda < 0.025$
for the \textit{bbh} data set of 1-OGC. 
In this plot, we use $m=10,429$ events whose $p$-value is less than 0.025. 
We have $0.98 < f(\lambda) <1.02$ for $0 < \lambda < 0.020$. 
We have a larger deviation from unity, for $0.020 < \lambda < 0.025$. 
This is due to the statistical fluctuation caused by smaller number in the numerator of Eq. (\ref{flambda}). 
From the same reason for the \textit{complete} data set, we set $\hat{\pi}_0=1$. 

Figure \ref{plot_f_lambda_2_all} is the plot of $f(\lambda)$ for the \textit{complete} data set of 2-OGC. 
In this plot, we use 103,185 events whose $p$-value is less than 0.07. 
We can see that $f(\lambda)$ in Eq. (\ref{flambda}) is almost unity for $0 < \lambda < 0.05$.
We have $0.995 < f(\lambda) <1.005$ in this region.  
Small deviation from unity near $\lambda = 0.07$ is due to the statistical fluctuation.

Figure \ref{plot_f_lambda_2_bbh} is the plot of $f(\lambda)$ for $0 < \lambda < 0.10$
for the \textit{bbh} data set of 2-OGC. 
In this plot, we use 152,759 events whose $p$-value is less than 0.10. 
We have $0.996 < f(\lambda) <1.002$ for $0 < \lambda < 0.08$. 
Larger deviation from unity for $0.020 < \lambda < 0.025$ is due to the statistical fluctuation. From the same reason as for 1-OGC data set, we set $\hat{\pi}_0=1$
in Step 3 of Algorithm ~\ref{alg2}.

\section{Derivation of Algorithm~\ref{alg2} to estimate $q$-values}
\label{sec:algorithm}

In this appendix, we discuss 
the estimation procedure of $q$-values.
We first introduce an algorithm which is a slight modification of
the procedure given in Remark B of Appendix of \cite{ST03}.
The input is the list of detection statistics obtained from the observed data and detection statistics in simulated background data.

\begin{algorithm}\label{alg1}

  Compute estimates of $q$-values defined in (\ref{q-value_est}).

\begin{itemize}
  
\item[1.] Compute estimated $p$-values
  \[
  \hat{p}_i\equiv \hat{p}(\rho_i)=\frac{n_{\rm bg}(\rho_i)}
    {n_{\rm bg}(0)},
  \]
  where $i = {1,...,n_{\rm bg}(0)}$, $\rho_i$ is the detection statistic of the $i$-th event and
  $n_{\rm bg}(\rho)$ is given in (\ref{nbg}).

\item[2.] Let $\hat{p}_{(1)}\le\hat{p}_{(2)}\le\cdots\le\hat{p}_{(n_{\rm obs})}$
  be the ordered $p$-values.

\item[3.] Set
  $\hat{q}_{(n_{\rm obs})}=\hat{\pi}_0\hat{p}_{(n_{\rm obs})}$.
  
\item[4.] For $i=n_{\rm obs}-1,n_{\rm obs}-2,...,1$, compute
  \[
    \hat{q}_{(i)}=
    \min\left(\frac{\hat{\pi}_0n_{\rm obs}\hat{p}_{(i)}}{i},\hat{q}_{(i+1)}
    \right). 
  \]

\item[5.] The estimated $q$-value for the $i$-th most significant event is
  $\hat{q}_{(i)}$ defined in (\ref{q-value_est}).

\end{itemize}
\end{algorithm}

Since Algorithm~\ref{alg1} is our starting point to construct Algorithm~\ref{alg2}, we reproduce it here.
If we set $m=n_{obs}$, Algorithm~\ref{alg2} reduces to Algorithm~\ref{alg1}.

Since $p$-values in the region around and larger than
$\hat{p}_{(m)}$ are noise, if we take the threshold
$u$ in $[\hat{p}_{(m)},1)$, 
we obtain $S(u) = n_1 + n_0 u$ where 
$n_0$ and $n_1$ are defined in Table \ref{table1}.
Accordingly, (\ref{FDRt_est}) is
\[
\widehat{\text{FDR}}(u)=\frac{\hat{\pi_0}
n_{\rm obs}u}{n_1+\hat{\pi_0}n_{\rm obs} u},
\]
which is monotonically increasing in $u$. Therefore, we  may replace Step 3 with
$\hat{q}_{(m)}=\hat{\pi}_0n_{\rm obs}\hat{p}_{(m)}/m$.
How Step 4
\begin{equation}
  \hat{q}_{(i)}
  =\text{min}\left(\frac{\hat{\pi}_0n_{\rm obs}\hat{p}_{(i)}}{i},
  \hat{q}_{(i+1)}\right)
\label{min}
\end{equation}
still gives (\ref{q-value_est}) for $i=m-1,m-2,...,1$ can be seen by induction.
Assume (\ref{min}) gives (\ref{q-value_est}) for $i=m-1,m-2,...,k+1$.
Note that
\begin{equation}
\hat{q}_{(m)}\ge\hat{q}_{(m-1)}\ge\cdots\ge\hat{q}_{(k+1)}\ge\hat{q}_{(k)}.
\label{mseq}
\end{equation}
Let us show that (\ref{min}) gives (\ref{q-value_est}) for $i=k$, namely,
\begin{equation}
  \frac{\hat{\pi}_0n_{\rm obs}u}
  {\#\{\hat{p}_{(j)}\le u;j\in\{1,...,m\}\}}
  \ge \hat{q}_{(k)}, \qquad \forall u\ge \hat{p}_{(k)}.
  \label{assertion}
\end{equation}
and the equality holds for some $u\ge\hat{p}_{(k)}$.

\begin{itemize}
  
\item If $\hat{\pi}_0n_{\rm obs}\hat{p}_{(k)}/k\le\hat{q}_{(k+1)}$, then
$\hat{q}_{(k)}=\hat{\pi}_0n_{\rm obs}\hat{p}_{(k)}/k$.
Note that the equality of (\ref{assertion}) holds if $u=\hat{p}_{(k)}$. For
$u\in(\hat{p}_{(k)},\hat{p}_{(k+1)})$,
\[
  \frac{\hat{\pi}_0n_{\rm obs}u}
  {\#\{\hat{p}_{(j)}\le u;j\in\{1,...,m\}\}}
  =\frac{\hat{\pi}_0n_{\rm obs}u}{k}
  >\hat{q}_{(k)}.
\]
For $u=\hat{p}_{(k+1)}$,
\[
  \frac{\hat{\pi}_0n_{\rm obs}u}
       {\#\{\hat{p}_{(j)}\le u;j\in\{1,...,m\}\}}
       =\frac{\hat{\pi}_0n_{\rm obs}\hat{p}_{(k+1)}}{k+1}
  \ge\hat{q}_{(k+1)}\ge \hat{q}_{(k)},
\]
where the second last inequality holds from (\ref{min}) and the last inequality
holds from (\ref{mseq}). Using the similar argument iteratively proves
the assertion.
  
\item If $\hat{\pi}_0n_{\rm obs}\hat{p}_{(k)}/k>\hat{q}_{(k+1)}$, then
$\hat{q}_{(k)}=\hat{q}_{(k+1)}$. For $u\in[\hat{p}_{(k)},\hat{p}_{(k+1)})$,
\[
  \frac{\hat{\pi}_0n_{\rm obs}u}
  {\#\{\hat{p}_{(j)}\le u;j\in\{1,...,m\}\}}
  =\frac{\hat{\pi}_0n_{\rm obs}u}{k}
  >\hat{q}_{(k+1)}=\hat{q}_{(k)}.
\]
For $u=\hat{p}_{(k+1)}$,
\begin{equation}
  \frac{\hat{\pi}_0n_{\rm obs}u}
       {\#\{\hat{p}_{(j)}\le u;j\in\{1,...,m\}\}}
       =\frac{\hat{\pi}_0n_{\rm obs}\hat{p}_{(k+1)}}{k+1}
       \ge\hat{q}_{(k+1)}= \hat{q}_{(k)}.
       \label{step}
\end{equation}
Suppose the second last equality holds, namely, the equality of
(\ref{assertion}) holds at $u=\hat{p}_{(k+1)}$. Then, for
$u\in(\hat{p}_{(k+1)},\hat{p}_{(k+2)})$,
\[
  \frac{\hat{\pi}_0n_{\rm obs}u}
  {\#\{\hat{p}_{(j)}\le u;j\in\{1,...,m\}\}}
  =\frac{\hat{\pi}_0n_{\rm obs}u}{k+1}
  >\hat{q}_{(k+1)}=\hat{q}_{(k)}.
\]
For $u=\hat{p}_{(k+2)}$,
\[
  \frac{\hat{\pi}_0n_{\rm obs}u}
  {\#\{\hat{p}_{(j)}\le u;j\in\{1,...,m\}\}}
  =\frac{\hat{\pi}_0n_{\rm obs}\hat{p}_{(k+2)}}{k+2}\ge
  \hat{q}_{(k+2)}\ge\hat{q}_{(k+1)}=\hat{q}_{(k)}.
\]
Using the similar argument iteratively proves the assertion.
If the second last equality of (\ref{step}) does not hold,
there exists some $l$ such that $k+2\le l\le m$ and
\[
       \frac{\hat{\pi}_0n_{\rm obs}\hat{p}_{(l)}}{l}
       =\hat{q}_{(l)}=\hat{q}_{(l-1)}=\cdots= \hat{q}_{(k)},
\]
because $\hat{q}_{(m)}=\hat{\pi}_0n_{\rm obs}\hat{p}_{(m)}/m$.
The assertion can be shown in the similar manner.

\end{itemize}

\bibliographystyle{ptephy}

\end{document}